\documentclass[aps,12pt,superscriptaddress,tightenlines,floatfix]{revtex4-1}
\usepackage{amsmath,amssymb,amsthm}
\usepackage[T1]{fontenc}
\usepackage[tt=false]{libertine}
\usepackage[scaled=0.83]{beramono}
\usepackage[libertine]{newtxmath}
\usepackage[dvips,dvipsnames]{xcolor}
\usepackage{graphicx}
\usepackage{bm}
\usepackage{natbib}

\newcommand{\nn}{\nonumber}

\newcommand{\R}{\mathbb{R}}
\renewcommand{\S}{\mathbb{S}}
\providecommand{\bra}[1]{\langle#1|}

\providecommand{\ket}[1]{|#1\rangle}

\DeclareMathOperator{\re}{Re}
\DeclareMathOperator{\im}{Im}
\DeclareMathOperator{\tr}{tr}
\providecommand{\abs}[1]{\lvert#1\rvert}
\providecommand{\smallabs}[1]{\lvert#1\rvert}

\providecommand{\norm}[1]{\lVert#1\rVert}
\providecommand{\smallnorm}[1]{\lVert#1\rVert}
\providecommand{\expectation}[1]{\mathsf{E}{#1}}
\providecommand{\median}[1]{\mathsf{M}{#1}}

\providecommand{\probability}[1]{\mathsf{P}{#1}}

\providecommand{\bra}[1]{\langle#1|}

\providecommand{\ket}[1]{|#1\rangle}

\providecommand{\operatorproduct}[3]{\langle#1\lvert#2\rvert#3\rangle}

\begin{document}
\title{Macroscopic Superposition States in Isolated Quantum Systems}
\author{Roman V. Buniy}
\email{roman.buniy@gmail.com}
\affiliation{Schmid College of Science, Chapman University, Orange, CA 92866}
\author{Stephen D.~H. Hsu}
\email{hsu@msu.edu}
\affiliation{Department of Physics and Astronomy, Michigan State University}
\date{\today}
\begin{abstract}
For any choice of initial state and weak assumptions about the Hamiltonian, large isolated quantum systems undergoing Schr\"{o}dinger evolution spend most of their time in macroscopic superposition states. The result follows from von Neumann's 1929 Quantum Ergodic Theorem. As a specific example, we consider a box containing a solid ball and some gas molecules. Regardless of the initial state, the system will evolve into a quantum superposition of states with the ball in macroscopically different positions. Thus, despite their seeming fragility, macroscopic superposition states are ubiquitous consequences of quantum evolution. We discuss the connection to many worlds quantum mechanics.
\end{abstract}
\maketitle

Highly complex superposition states have been realized in the laboratory \cite{Nobel}. Despite their seeming fragility, such states are of great importance in quantum information and computing, as well as to theoretical questions in quantum foundations. It may come as a surprise that isolated systems with many degrees of freedom naturally evolve into macroscopic superposition states. Such states contain orthogonal components which differ in macroscopic quantities such as the position or momentum of large objects normally thought to be ``classical'' in nature. 

In what follows we use a specific example (essentially, that of Brownian motion) to illustrate how this result follows from the 1929 Quantum Ergodic Theorem (QET) of John von Neumann \cite{QET}. This theorem was largely forgotten for over 50 years, until its resurrection in 2009-10 \cite{GLMTZ,GLTZ}. The QET contains insights relevant both to quantum statistical mechanics and the foundations of quantum mechanics. We conclude with some remarks on the latter topic.

The QET goes beyond typicality (concentration of measure) results concerning thermalization in isolated quantum systems \cite{typical}. The typicality result says that almost all pure states $\Psi$ of large systems are maximally entangled, and tracing over all but a small subspace $1$ yields a density matrix $\rho_1$ which is close to the normalized identity, i.e. the microcanonical density matrix. This implies thermal properties for the small subspace. The QET is focused specifically on the subspace of macroscopic observables, as opposed to a generic subset of microscopic degrees of freedom. Von Neumann proves a stronger result concerning time evolution (ergodicity) of the system: that \textit{all} initial states $\Psi_0$ will spend most of their time evolution as typical states with respect to the macro subspace (see equation \eqref{QET} below), subject of course to certain assumptions required for the theorem \cite{RS}.

The calculation given below illustrates that for any subspace of a large system (including, for example, a subspace defined by a set of macroscopic observables), the density operator $\rho_1$ induced by tracing over the other degrees of freedom of a random pure state is overwhelmingly likely to be close to $\rho_1 \approx n_1^{-1} I_1$. Based on this dominance of the measure, one can argue heuristically that even if the system begins in an exceptional state where this property is strongly violated, dynamical evolution will cause it to spend most of its time in a typical state. The QET provides a rigorous basis for this intuition.

Let $\{\phi_{1,j_1}\}_{j_1=1}^{n_1}$ and $\{\phi_{2,j_2}\}_{j_2=1}^{n_2}$ be complete sets of orthonormal eigenstates of two Hilbert spaces labeled 1 and 2, with identity operators $I_1$ and $I_2$.
For a state
\begin{align}
  \psi =\sum_{j_1=1}^{n_1} \sum_{j_2=1}^{n_2}  c_{j_1,j_2} \phi_{1,j_1}\otimes \phi_{2,j_2}
  \label{}
\end{align}
the corresponding density operator $\rho=\psi\psi^\dagger$ and the partial density operator $\rho_1=\tr_2{\rho}$ are
\begin{align}
  \rho &=\sum_{j_1,k_1=1}^{n_1} \sum_{j_2,k_2=1}^{n_2} c_{j_1,j_2} c^*_{k_1,k_2} \phi_{1,j_1} \phi_{1,k_1}^\dagger \otimes\phi_{2,j_2} \phi_{2,k_2}^\dagger, \\
  \rho_1 &=\sum_{j_1,k_1=1}^{n_1} \sum_{j_2=1}^{n_2} c_{j_1,j_2} c^*_{k_1,j_2} \phi_{1,j_1} \phi_{1,k_1}^\dagger
  \label{}
\end{align}
We assume that $\sum_{j_1=1}^{n_1}\sum_{j_2=1}^{n_2}\smallabs{c_{j_1,j_2}}^2=1$, in which case $\smallnorm{\psi}^2=1$ and $\tr{\rho}=\tr_1{\rho_1}=1$.
For the uniform distribution on $S^{2n_1 n_2-1}$ we have the expected value
\begin{align}
  \expectation{(c_{j_1,j_2} c^*_{k_1,k_2})} &=(n_1 n_2)^{-1}\delta_{j_1,k_1}\delta_{j_2,k_2}, \label{expectation_2} 
\end{align}
which implies $\expectation{\rho_1} =n_1^{-1} I_1$.

We can show that $\rho_1$ is strongly concentrated near its expected value, with small probability of deviation, $\probability{}$. We use Levy's lemma \cite{Levy,Milman,Jenkinson}, which implies that for any Lipschitz continuous function $f \colon \S^{2n_1 n_2-1} \to \R$, a uniformly distributed random variable $c\in  \S^{2n_1 n_2-1}$ and any $\epsilon>\delta$, 
\begin{align}
  \probability{\bigl(\abs{f(c)-\expectation{f(c)}}\ge\epsilon\bigr)} \le  \exp{\biggl(- \frac{n_1 n_2 (\epsilon - \delta)^2}{\norm{f}_{\textrm{L}}^2}\biggr)},
  \label{Levy}
\end{align}
where
\begin{align}
  \norm{f}_{\textrm{L}}=\sup_{c\in\S^{2 n_1 n_2-1}}\norm{\nabla_c f(c)}
  \label{}
\end{align}
is the Lipschitz norm of $f$ and $\delta = ( \frac{\pi}{4 n_1 n_2} )^{1/2}$. The specific bound \eqref{Levy} is deduced using Proposition 1.8 in \cite{Levy}. (See Appendix for additional details.) The result is a consequence of the extreme concentration of measure exhibited in the geometry of high dimensional spheres \cite{Levy,Milman}. Given that $\rho_1$ is quadratic in $c$, which is itself bounded in norm, we can bound the gradient $\nabla_c \rho_1$ to obtain an exponential concentration in probability: 
\begin{align}
  & \probability{\bigl(\smallabs{\re{(\rho_1)_{j_1,j_1}} -n_1^{-1}} \ge \epsilon \bigr)} \le  \exp{ \bigl( - \tfrac{n_1 n_2}{4} (\epsilon - \delta)^2} \bigr), \label{probability_rediag_rho_1_levy} \\
  & \probability{\bigl(\smallabs{\re{(\rho_1)_{j_1,k_1}}} \ge \epsilon \bigr)} \le \exp{\bigl(- n_1 n_2 \epsilon^2} \bigr), \label{probability_re_rho_1_levy} \quad j_1\not=k_1, \\
  & \probability{\bigl(\smallabs{\im{(\rho_1)_{j_1,k_1}}} \ge \epsilon \bigr)} \le \exp{\bigl(-n_1 n_2 \epsilon^2 \bigr)}, \quad j_1\not=k_1. \label{probability_im_rho_1_levy}
\end{align}
In particular, as $n_1 n_2$ becomes large the density matrix $\rho_1$ becomes exponentially concentrated near 
$\rho_1 =n_1^{-1} I_1$. The set of states $\psi$ for which $\rho_1$ deviates from this value approaches measure zero. (Note neither $n_1$ nor $n_2$ has to be individually large, as long as their product is.) Thus it is plausible, and the QET demonstrates, that under dynamical evolution of $\psi$: $\rho_1 \approx n_1^{-1} I_1$ almost all of the time.

Now let us turn to a specific example, and focus on macroscopic superposition states.
Consider a solid ball and gas molecules isolated in a box. The Hamiltonian can be divided into three terms: $H_{\textrm{B}}$ describing the ball, $H_{\textrm{G}}$ for the gas molecules, and $H_{\textrm{BG}}$ describing interactions between the two. Let 
\begin{equation}
H_{\textrm{B}} = \frac{P^2}{2M} + V (Q) + \sum_{i=1}^{N} \frac{p_i^2}{2m} +  U(q_1,\dotsc, q_N ),
\end{equation}
where $P$ and $Q$ are center of mass momentum and position, and $i$ label relative coordinates and momenta \cite{Messiah} of the $N+1$ individual molecules making up the ball. Importantly, $P$ and $Q$ commute with the relative degrees of freedom.

We take the potential $V$ to be non-zero, so that the energy of the ball varies very slightly with location in the box (i.e., the bottom of the box is not exactly flat). This assumption is motivated by technical aspects of the proof of the QET (specifically the assumption that energy levels and gaps are non-degenerate) but does not affect the main physical aspects of our discussion. In a realistic setting we expect that the technical assumptions used in the QET to be satisfied. The details of $H_{\textrm{G}}$ and $H_{\textrm{BG}}$ are not important, except that we assume that the gas molecules are excluded from the space occupied by the ball. For example, $H_{\textrm{BG}}$ may contain repulsive potential terms which depend on $Q$.

The macroscopic quantities we are interested in are the total energy of the system and the center of mass position $Q$ and momentum $P$ of the ball. The macroscopic energy operator is a coarse grained version of the (microscopic) Hamiltonian $H = H_{\textrm{B}} + H_{\textrm{G}} + H_{\textrm{BG}}$. The eigenvalues of $H$ are grouped into bands (energy shells) whose width is macroscopic (i.e., there are many eigenvalues in each band; the energy difference between bands is experimentally accessible). Each of the $H$ eigenstates in a band is assigned a coarse grained eigenvalue equal to the average energy in the band, so that the macroscopic energy operator has very large degeneracies (all of the states in a band share the same eigenvalue).

The definition of the macroscopic energy given above can be generalized: given a microscopic operator $A$, which has eigenstate expansion $A = \sum_a a \ket{a}\bra{a}$, the coarse grained or macroscopic version is obtained by dividing the sum into bands of nearby eigenvalues, and replacing the coefficients $a$ with the average within the band.

An important property of macroscopic observables is that they can be measured simultaneously, so the corresponding operators must commute. The macroscopic momentum $P$ and position $Q$ of the ball are constructed using wave packet states which have simultaneously a central value of momentum and of position, and a spread in each which is small but respects the uncertainty principle. These states are orthogonalized, resulting in simultaneous eigenstates for macroscopic $P$ and $Q$. The procedure is discussed in von Neumann's proof of the QET.

The full set of mutually commuting macroscopic operators $\{ M_i \}$ includes coarse grained versions of $H, P, Q$. Within a specific energy shell of fixed macroscopic energy there are many individual phase space cells with different values of macroscopic $(P,Q)$. In the general case other quantities may be included in $\{ M_i \}$.

We restrict attention to a specific energy shell of dimension $D$, which is further partitioned into subspaces with definite values of each of these macroscopic variables $\{ M_i \}$. Let $\nu$ denote a full set of values for the macroscopic variables, and $d_\nu$ denote the dimensionality of the subspace with values labeled by $\nu$, so $D = \sum_\nu d_\nu$.

The Quantum Ergodic Theorem \cite{QET,GLMTZ,GLTZ} states that: for any initial state $\Psi_0$ and almost any Hamiltonian $H$, in the long run the system spends almost all of its time in typical states $\Psi$ with the property that
\begin{align}
\label{QET}
\operatorproduct{\Psi}{\bm{P}_\nu}{\Psi} \approx \frac{d_\nu}{D},
\end{align}
where $\bm{P}_\nu$ is a projector onto the macro subspace labeled by $\nu$. The conditions on $H$ are the absence of degeneracies (which are unlikely, in the absence of exact symmetries) and a technical condition on the relation between the energy eigenstate basis and the $\nu$ basis. In the limit of large number of degrees of freedom the technical condition is violated only for a vanishingly small subset of macroscopic operators \cite{Reimann}.  Both conditions can be made to hold in realistic settings, such as the case of Brownian motion of a macroscopic ball interacting with a gas, for natural choices of $H$. In particular, if a specific setup  does not satisfy the conditions (e.g., choice of interaction terms, box geometry, or specific coarse grained macroscopic operator; this would be highly unlikely for the reasons given), a slightly perturbed version of the setup is almost certain to suffice. The theorem has been explicitly tested in simulations of specific systems \cite{RS}.

The QET implies that in the typical state (i.e., most of the time, in the long run) there is significant probability to find the ball in any of the phase space cells labeled by $\nu$, as long as the $d_\nu$ are similar in magnitude. In our example this would mean that the ball position could be roughly anywhere in the box. 

We can show that {\it this corresponds to a macroscopic superposition state} by noting that $\Psi$ must be a pure state at all times, which by the Schmidt decomposition must have the form
\begin{align}
\Psi = \sum_\nu c_\nu \psi_\nu \otimes \phi_\nu  
\end{align}
where $\psi_{\nu}$ describes the (macroscopic) state of the ball and $\phi_{\nu}$ describes the gas molecules. By construction, the $\nu$ states are mutually orthogonal: i.e., $\psi_\nu$ has zero overlap with $\psi_{\nu'}$ for $\nu \neq \nu'$ and similarly with the $\phi$ states. The gas molecule (``environment'') state is entangled with the ball state, as the gas molecules cannot occupy the same space as the ball. The QET implies that the magnitudes $\abs{c_\nu}$ are comparable as long as the corresponding $d_\nu$ are similar in size. In concrete terms this assumption is satisfied for $\nu$ corresponding to phase space cells $(P,Q)$ in which the macroscopic energies of the ball are nearly the same, leaving similar volumes of phase space to be occupied by the gas molecules  (subject to the total energy constraint for the given macroscopic energy shell).

Note the claim is \textit{not} that the ball spends some of the time at $\nu$ and some at $\nu'$. Our focus is not the time average of its position. The claim is stronger: that at almost all times the ball is in a macroscopic superposition state, and the entangled environment (air molecules, or even an observer whose brain is a macroscopic neural net) is as well. This all follows from the assumption of unitary (Schr\"{o}dinger) evolution inside the box, using the QET. The basis denoted by $\nu$ is the \textit{natural basis} for a macroscopic observer of the system.

We can give the following intuitive (quasi-classical) explanation for the result. The ball is subject to fluctuations arising from collisions with the gas --- i.e., Brownian motion. Over long enough timescales there is a non-negligible amplitude for fluctuations to cause the ball to random walk from its starting position to any other position in the box. As Schr\"{o}dinger dynamics is linear, $\Psi$ evolves into a superposition of the possible outcome states. The location of the ball is reflected in the corresponding state of the gas molecules (ball and gas cannot occupy the same space), which act as an environment for the ball. After enough time, the total wavefunction can be decomposed into orthogonal branches which differ macroscopically. This description suggests that the timescale over which an initially localized ball state evolves into a typical $\Psi$ satisfying equation (\ref{QET}) is similar to that for a Brownian random walk to reach all parts of the box.

The example above can be generalized to almost any system whose dynamics permit multiple macroscopic outcomes. The QET implies that the quantum state evolves into a superposition over these outcomes. The ubiquity of macroscopic superposition states under ordinary Schr\"{o}dinger evolution is of course an aspect of many worlds, or no collapse, quantum mechanics \cite{manyworlds}. This view of quantum mechanics is typically introduced via the measurement process, which we briefly summarize below.

Let $Q$ be a single qubit and $M$ a macroscopic device which measures the spin of the qubit along a particular axis. The eigenstates of spin along this axis are denoted $\vert \pm \rangle$. We define the operation of $M$ as follows, where the combined system is $S = Q + M$:
\begin{align}
&\ket{+}\otimes\ket{M} \longrightarrow \ket{S_{+}}, \\
&\ket{-}\otimes\ket{M} \longrightarrow \ket{S_{-}},
\end{align}
where $S_+$ denotes a state of the total system in which the apparatus $M$ has recorded a $+$ outcome (i.e., is in state $M_+$), and similarly with $S_-$. We can then ask what happens to a superposition state 
$\ket{\Psi_Q} = c_+ \ket{+} + c_- \ket{-}$ which enters the device $M$. In the conventional formulation, with measurement collapse, {\it one} of the two final states $\ket{S_+}$ {\it or} $\, \ket{S_-}$ is realized, with probabilities $\vert c_+ \vert^2$ and $\vert c_- \vert^2$ respectively. 

However, if the combined system $S = Q + M$ evolves according to the Schr\"{o}dinger equation (in particular, linearly), we obtain a superposition of measurement device states:
\begin{align}
\big(  c_+ \ket{+} ~+~ c_- \ket{-}  \big ) \otimes \ket{M} ~~ \longrightarrow ~~
c_+ \, \ket{S_+} ~+~ c_- \, \ket{S_-}.
\end{align}
This seems counter to actual experience: measurements produce a single outcome, not a superposition state. But an observer in the state $S_+$ might be unaware of the second branch of the wave function in state $S_-$ for dynamical reasons related to a phenomenon called {\it decoherence} \cite{decoherence}. Any object sufficiently complex to be considered either a measuring device or observer (i.e., which, in the conventional formulation can be regarded as semi-classical) will have many degrees of freedom. A measurement can only be said to have occurred if the states $M_+$ and $M_-$ are very different: the outcome of the measurement must be stored in a redundant and macroscopically accessible way in the device (or, equivalently, in the local environment). Therefore, the overlap of $M_+$ with $M_-$ is of order $\exp( - N )$, where $N$ is a macroscopic number of degrees of freedom, and the future dynamical evolution of each branch is unlikely to alter this. For all practical purposes, as John Bell put it \cite{Bell}, an observer on one branch can ignore the existence of the other: they are said to have decohered. Each of the two observers will perceive a collapse to have occurred, although the evolution of the overall system $S$ has continued to obey the Schr\"{o}dinger equation. 

\smallskip

The Quantum Ergodic Theorem shows that macroscopic superpositions of outcome states like $\ket{S_\pm}$ are ubiquitous under Schr\"{o}dinger evolution. Interactions such as the scattering of particles in Brownian motion lead to such states, even if observers in the closed system (described by the total wave function $\Psi$) are unaware of it. 

\appendix
\section{}

Writing $c_{j_1,j_2}=c'_{j_1,j_2}+ic''_{j_1,j_2}$, where $c'_{j_1,k_1}, c''_{j_1,k_1}\in \R$,
\begin{align}
  \sum_{j_1=1}^{n_1} \sum_{j_2=1}^{n_2} (c^{\prime 2}_{1,j_1,k_1} +c^{\prime\prime 2}_{j_1,k_1}) =1,
  \label{}
\end{align}
and using
\begin{align}
  &\re{(\rho_1)_{j_1,k_1}} =\sum_{j_2=1}^{n_2} (c'_{j_1,j_2}c'_{k_1,j_2} +c''_{j_1,j_2}c''_{k_1,j_2}), \\
  &\im{(\rho_1)_{j_1,k_1}} =\sum_{j_2=1}^{n_2} (c''_{j_1,j_2}c'_{k_1,j_2} -c'_{j_1,j_2}c''_{k_1,j_2}),
  \label{}
\end{align}
we find
\begin{align}
  \norm{\nabla_c \re{(\rho_1)_{j_1,k_1}}}^2 &=\norm{\bigl( \delta_{l_1,j_1} c'_{k_1,l_2} +c'_{j_1,l_2}\delta_{l_1,k_1}, \delta_{l_1,j_1} c''_{k_1,l_2} +c''_{j_1,l_2}\delta_{l_1,k_1} \bigr)_{1\le l_1\le n_1,1\le l_2\le n_2}}^2 \nn \\
  &=\sum_{l_2=1}^{n_2} \bigl( c^{\prime 2}_{j_1,l_2} +c^{\prime \prime 2}_{j_1,l_2} +c^{\prime 2}_{k_1,l_2} +c^{\prime\prime 2}_{k_1,l_2} \bigr) +2\delta_{j_1,k_1}\sum_{l_2=1}^{n_2} \bigl( c^{\prime 2}_{j_1,l_2} +c^{\prime\prime 2}_{j_1,l_2} \bigr) \nn \\
  &=\sum_{l_2=1}^{n_2} \bigl( \smallabs{c_{j_1,l_2}}^2 +\smallabs{c_{k_1,l_2}}^2 \bigr) +2\delta_{j_1,k_1}\sum_{l_2=1}^{n_2} \smallabs{c_{j_1,l_2}}^2 \label{} \\
  \norm{\nabla_c \im{(\rho_1)_{j_1,k_1}}}^2 &=\norm{\bigl( c''_{j_1,l_2}\delta_{l_1,k_1} -\delta_{l_1,j_1}c''_{k_1,l_2}, \delta_{l_1,j_1}c'_{k_1,l_2} -c'_{j_1,l_2}\delta_{l_1,k_1} \bigr)_{1\le l_1\le n_1,1\le l_2\le n_2}}^2 \nn \\
  &=\sum_{l_2=1}^{n_2} \bigl( c^{\prime 2}_{j_1,l_2} +c^{\prime\prime 2}_{j_1,l_2} +c^{\prime 2}_{k_1,l_2} +c^{\prime\prime 2}_{k_1,l_2} \bigr) -2\delta_{j_1,k_1}\sum_{l_2=1}^{n_2} \bigl( c^{\prime 2}_{j_1,l_2} +c^{\prime\prime 2}_{j_1,l_2} \bigr) \nn \\
  &=\sum_{l_2=1}^{n_2} \bigl( \smallabs{c_{j_1,l_2}}^2 +\smallabs{c_{k_1,l_2}}^2 \bigr) -2\delta_{j_1,k_1}\sum_{l_2=1}^{n_2} \smallabs{c_{j_1,l_2}}^2. \label{}
\end{align}
As a result,
\begin{align}
  &\norm{\nabla_c \re{(\rho_1)_{j_1,j_1}}}^2 \le 4, \label{} \\
  &\norm{\nabla_c \re{(\rho_1)_{j_1,k_1}}}^2 \le 1, \quad j_1\not=k_1, \label{} \\
  &\norm{\nabla_c \im{(\rho_1)_{j_1,k_1}}}^2 \le 1, \quad j_1\not=k_1. \label{}
\end{align}
For bounds \eqref{probability_re_rho_1_levy} and \eqref{probability_im_rho_1_levy} on the off-diagonal terms in $\rho_1$ we use a somewhat stronger inequality as the median $\median{f}$ and expectation $\expectation{f}$ coincide: $\probability{(\abs{f- \median{f}}\ge \epsilon)} \le \exp ( - n_1 n_2 \epsilon^2 \norm{f}^{-2}_{\textrm{L}} )$, Proposition A.0.5 in \cite{Jenkinson}.
For probability of deviation from expectation we use Proposition 1.8 in \cite{Levy} and the median result to obtain
\begin{align}
  \probability{(\abs{f-\expectation{f}}\ge \epsilon)} \le \exp{\Bigl(-\frac{n_1 n_2 (\epsilon-\delta)^2}{\norm{f}^2_{\textrm{L}}}\Bigr)}
  \label{P1.8}
\end{align}
with $\delta = (\frac{\pi}{4 n_1 n_2})^{1/2} \norm{f}_{\textrm{L}}$ and $\epsilon \ge \delta$. 
The specific bounds we obtain from concentration of measure are stronger than those previously given in the literature as expressions of Levy's Lemma. They can be applied, for example, to relatively small quantum systems in the laboratory.


\end{document}